\documentclass[%
reprint,
superscriptaddress,
 amsmath,amssymb,
 aps,
prl,
]{revtex4-2}

\usepackage{graphicx}
\usepackage{dcolumn}
\usepackage{bm}
\usepackage{float}

\begin{document}


\title{Spontaneous voltage peaks in superconducting Nb channels without engineered asymmetry}

\author{Shamashis Sengupta}
\email[]{shamashis.sengupta@ijclab.in2p3.fr}
\affiliation{Universit\'{e} Paris-Saclay, CNRS/IN2P3, IJCLab, 91405 Orsay, France}

\author{Miguel Monteverde}
\affiliation{Universit\'e Paris-Saclay, CNRS, Laboratoire de Physique des Solides, 91405, Orsay, France}

\author{Sara Loucif}
\affiliation{Universit\'{e} Paris-Saclay, CNRS/IN2P3, IJCLab, 91405 Orsay, France}

\author{Florian Pallier}
\affiliation{Universit\'{e} Paris-Saclay, CNRS/IN2P3, IJCLab, 91405 Orsay, France}

\author{Louis Dumoulin}
\affiliation{Universit\'{e} Paris-Saclay, CNRS/IN2P3, IJCLab, 91405 Orsay, France}

\author{Claire Marrache-Kikuchi}
\affiliation{Universit\'{e} Paris-Saclay, CNRS/IN2P3, IJCLab, 91405 Orsay, France}


\begin{abstract}
Rectification effects in solid-state devices are a consequence of nonreciprocal transport properties. This phenomenon is usually observed in systems with broken inversion symmetry. In most instances, nonreciprocal transport arises in the presence of an applied magnetic field and the rectified signal has an antisymmetric dependence on the field. We have observed rectification of environmental electromagnetic fluctuations in plain Nb channels without any asymmetry in design, leading to spontaneous voltage peaks at the superconducting transition. The signal is symmetric in the magnetic field and appears even without an applied field at the critical temperature. This is indicative of an unconventional mechanism of nonreciprocal transport resulting from a spontaneous breaking of inversion symmetry.
\end{abstract}

\maketitle



Rectification of a.c. signals into a d.c. voltage is known in superconducting devices incorporating some form of asymmetry in structure, either at atomic or mesoscopic lengthscales. Rectification effects have their origin in nonreciprocal transport properties induced by a breaking of inversion symmetry. In most cases, a breaking of time reversal symmetry by application of a magnetic field is also required. Nonreciprocal transport in an applied magnetic field has been observed in non-centrosymmetric superconductors \cite{wakatsuki, zhang}, polar superconductors with Rashba-type spin-orbit interaction \cite{itahashi}, topological insulator/superconductor interfaces \cite{yasuda} and amorphous superconductors attached to a magnetic material \cite{lustikova}. In devices employing the principle of vortex ratchets, the asymmetry is engineered through artificially created pinning arrays \cite{villegas, silva, vondel}. The lack of inversion symmetry in certain types of superconducting devices \cite{ando, bauriedl, baumgartner, wu} leads to the observation of different critical currents for opposite bias polarities. This phenomenon, called the superconducting diode effect, is presently a very active area of research. Some  systems showing nonreciprocal properties are capable of rectifying environmental electromagnetic fluctuations into a d.c. voltage. This has been observed in MoGe (superconductor)/Y$_3$Fe$_5$O$_{12}$ (insulator magnet) heterostructures \cite{lustikova} and thin flakes of NbSe$_2$ \cite{zhang}. A common feature of such nonreciprocal transport, induced by an applied magnetic field, is that the observed voltage has an antisymmetric field-dependence \cite{wakatsuki, zhang, itahashi, yasuda, lustikova, ando, wu, cerbu, pryadun}. In this work, we report the observation of spontaneous voltage peaks in plain channels of superconducting Nb without any asymmetry in its design. The voltage peaks are symmetric with respect to the magnetic field. These features are observed even without the application of a magnetic field, at the critical temperature of the transition. This highlights an unconventional mechanism of rectification in a superconducting device which is different from the most common examples of this phenomenon.

The experiments were conducted on superconducting channels of Nb in various geometries, realized on commercially available Si wafers covered with 500 nm of SiO$_2$. The devices were patterned using electron beam lithography, following which we evaporated a layer of Nb (thickness between 55 nm to 72 nm). The superconducting critical temperature ($T_c$) of the samples varied from one batch to another. The $T_c$ of niobium depends strongly on the quality of vacuum in the deposition chamber \cite{desorbo}  and average grain size \cite{bose}. Further, contamination from electron beam resist contributed to a significantly lower $T_c$ of narrow channels with widths below a few micrometres. We were thus able to study superconducting devices with a wide range of critical temperatures (2.5 K to 7.7 K). Electrical contacts were established with ultrasonic wire-bonding directly on contact pads of the sample. The electrical lines inside the cryostat used for transport measurements have a resistance of about 1 $\Omega$ each, measured from the external connectors at room temperature down to the sample stage. 

Fig. 1a shows the schematic diagram of a sample (named D1) with a simple geometry. It consists of a central channel of width ($w$) 38 $\mu$m and length ($L$) 245 $\mu$m, connected to two large contact pads. The thickness of the Nb film is 72 nm. (See Supplemental Material for images.) The variation of resistance ($R$) measured in two-probe configuration with temperature ($T$) is shown in Fig. 1b. We determine $T_c$ and the upper critical magnetic field ($B_{c2}$) by the midpoint of the resistance drop. The direction of applied magnetic field was always perpendicular to the plane of the sample.

There are two superconducting transitions in Fig. 1b, corresponding to the critical temperatures of the large pads (5.1 K) and the central channel (3.9 K). At a temperature of 2.5 K, we observed $B_{c2}$ of 2.8 T for the pads and 1.6 T for the channel (Fig. 1c). We will now focus on the response of the device in the absence of any applied bias current. For this, we simply connected a voltmeter across the sample. Fig. 1d shows a schematic diagram of this measurement. The two contacts pads on the device are labelled T1 and T2. There are two ways of connecting the voltmeter leads across a two-terminal device. These are marked as Configurations 1 and 2. We adopt the convention of denoting the measured voltage as V[T1,T2], where the first (second) contact is the one connected to the positive (negative) terminal of the voltmeter input. The voltmeter used for this particular measurement was a Keithley DMM6500 Multimeter (henceforth DMM). Upon sweeping the magnetic field, we observed two sharp peaks in the measured voltage (Fig. 1e) almost coinciding with $B_{c2}$ of the central channel. These spontaneously arising peaks are not transient features. The measured voltage is steady in time if the magnetic field is held constant.

\begin{figure}
\begin{center}
\includegraphics[width=88mm]{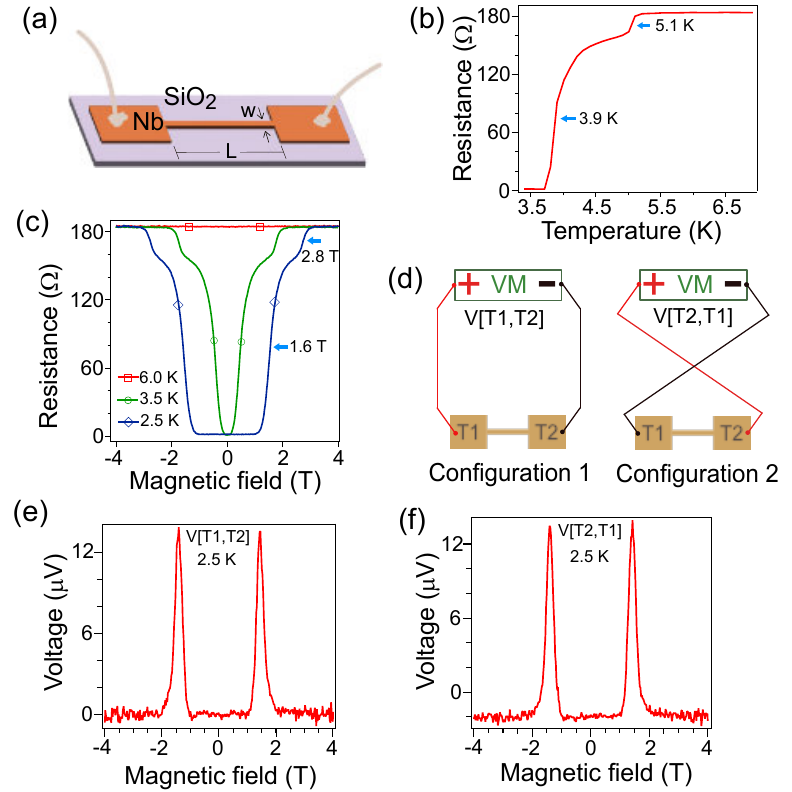}
\caption{\textbf{(a)} Representation of the geometry of sample D1. \textbf{(b, c)} Observation of the superconducting transition by measuring the resistance as a function of temperature (b) and magnetic field (c). The d.c. current applied was 1 $\mu$A. \textbf{(d)} Schematic diagram of the measurement of voltage across the sample without any applied bias current using a d.c. voltmeter (VM). \textbf{(e)} Spontaneous voltage peaks as a function of magnetic field with the circuit of Configuration 1. \textbf{(f)} Spontaneous voltage peaks as a function of magnetic field with the circuit of Configuration 2.}
\end{center}
\end{figure}

\begin{figure*}[ht]
\begin{center}
\includegraphics[width=184mm]{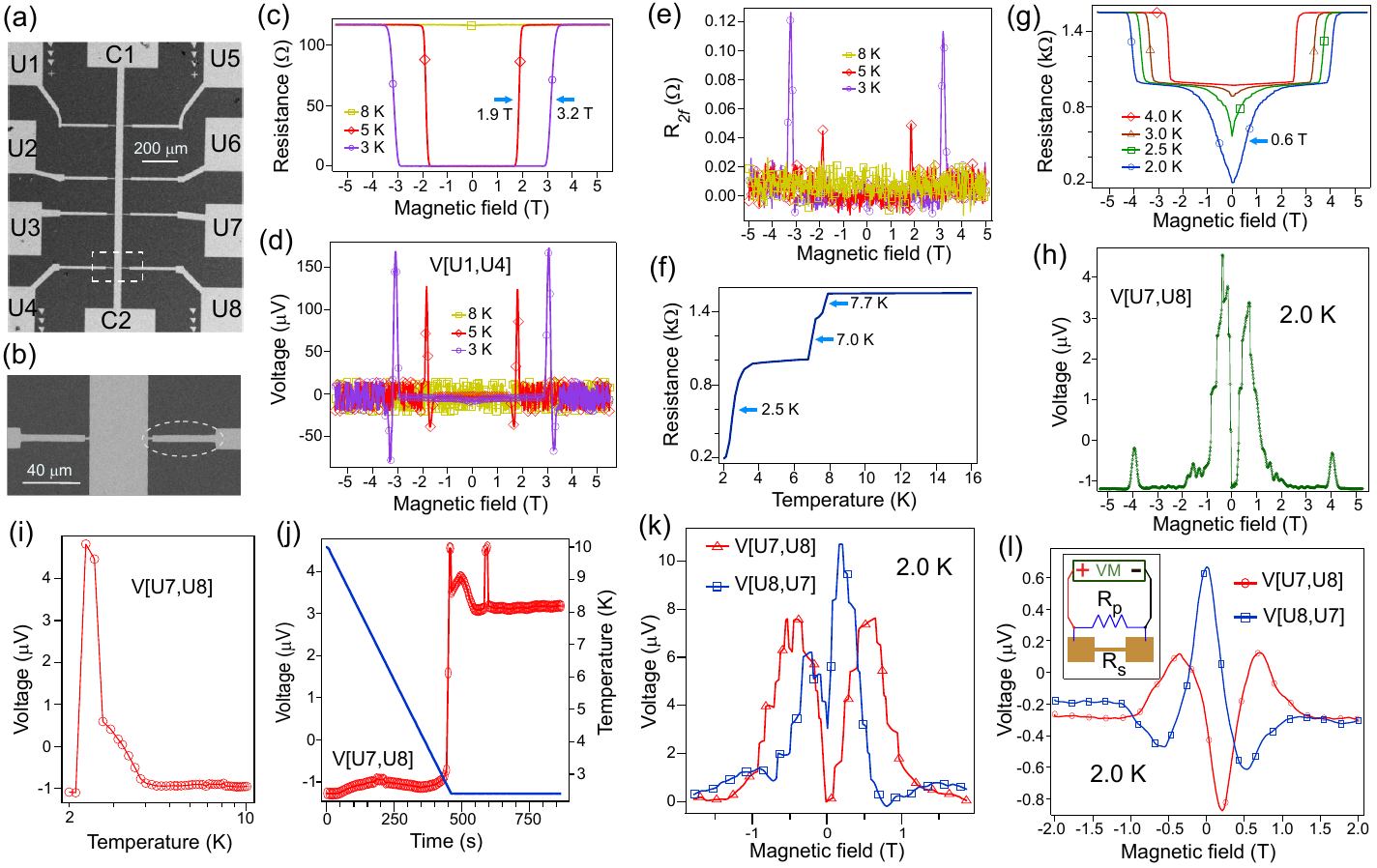}
\caption{\textbf{(a,b)} Image of sample D2 in the scanning electron microscope (SEM). A close-up of the narrow channels of the voltage probes (dashed rectangle in (a)) is shown in (b). \textbf{(c)} Four-probe measurement of resistance using a low frequency lock-in technique. \textbf{(d)} Measurement of V[U1,U4] with a DMM simultaneously with the resistance measurement described in (c). \textbf{(e)} $R_{2f}$ estimated from the voltage at the second harmonic frequency measured between U1 and U4. \textbf{(f,g)} Two-probe resistance measurement as a function of temperature (f) and magnetic field (g) across U7 and U8, using an applied d.c. current of 78 nA. \textbf{(h)} Measurement of V[U7,U8] without any applied bias current with a nanovoltmeter as a function of magnetic field. \textbf{(i)} Measurement of V[U7,U8] without any applied bias current with a nanovoltmeter as a function of temperature at zero field. \textbf{(j)} The spontaneously arising voltage V[U7,U8] (open circles) is measured with a variation of temperature (line). When the temperature reaches 2.4 K, it is held constant and the voltage is recorded with time. \textbf{(k)} Measurement of spontaneous voltage across U7 and U8 as a function of magnetic field by interchanging the polarities of nanovoltmeter leads on the two contacts. \textbf{(l)} Measurement of V[U7,U8] and V[U8,U7] as a function of the magnetic field. (Inset) Circuit implemented with a resistance $R_p$ in parallel with the sample.}
\end{center}
\end{figure*}

The measurement of a finite voltage by a voltmeter implies that a current is flowing through its internal resistance. Therefore, power is continuously dissipated when the voltage is steady in time. We can obtain insights about the physical origin of the spontaneous voltage peaks (Fig. 1e) by identifying the source of energy. Heat flowing to the sample from the thermal bath (sample stage) cannot be considered as a possibility. This would imply that heat extracted from a source is converted completely into electrical work, in violation of the second law of thermodynamics. A more reasonable explanation is the rectification of environmental electromagnetic fluctuations by the device at the superconducting transition. This phenomenon, that a superconducting system acts as a rectifying antenna capable of generating d.c. electricity from nonequilibrium environmental electromagnetic fluctuations, has been observed in  MoGe/Y$_3$Fe$_5$O$_{12}$ heterostructures \cite{lustikova} and NbSe$_2$ \cite{zhang}. The voltage peaks at the critical field in our devices show clear similarities with observations reported in these works \cite{foot}. The surprising fact, however, is that we observe a rectification effect at all, since there is no apparent source of asymmetry in our devices. In Refs. 2 and 5, the voltage peaks are antisymmetric as a function of the magnetic field. The symmetric-in-field peaks in our experiments highlight that a different mechanism of rectification is at work here.

A d.c. voltage requires a preferential direction to determine its polarity. How this biasing happens is a key question that needs to be understood. If the polarities develop at random, an even distribution of positive and negative peaks is expected. However, in most instances, we observed positive peaks in our measurements (for reasons that will be explained later). Further cross-checks yielded unexpected results. We interchanged the leads of the voltmeter on the contact pads to measure V[T2,T1] (Configuration 2 in Fig. 1d). We would expect V[T2,T1] to have opposite sign of V[T1,T2]. However, upon sweeping the magnetic field, we still observed (Fig. 1f) positive voltage peaks. This shows that features of the spontaneous d.c. voltage are not only related to the superconducting transition in the sample, but also to an influence of the internal circuitry of the voltmeter. Further discussions on this issue will be presented later.

We now present results from a second sample D2 (Fig. 2a). This consisted of a central channel ($w$ = 40 $\mu$m) for applying a current (between contacts C1 and C2) and  eight probes for measuring voltage (numbered from U1 to U8). The thickness of Nb film was 55 nm. The voltage probes consisted of narrow channels a few micrometres wide (Fig. 2b), with reduced $T_c$ ($<$ 3.0 K). Standard four-probe measurements showed a sharp superconducting transition at 7.7 K in the central channel connecting C1 and C2. (See Supplemental Material for the characterization of sample D2.) Fig. 2c shows the result of four-probe resistance measurement using a standard lock-in technique. A current of 1 $\mu$A r.m.s. was applied between C1 and C2. The voltage was measured between the probes U1 and U4 with a Stanford Research Systems SR830 lock-in amplifier. Simultaneously, a DMM measured the d.c. voltage between the same contacts U1 and U4 (Fig. 2d). The resistance drop at the superconducting transition at $B_{c2}$ (Fig. 2c) was accompanied by d.c. voltage peaks (Fig. 2d).

The resistance of a system exhibiting nonreciprocal properties, as a function of current ($I$), may be written as:
\begin{equation}
R=R_0+\alpha I
\end{equation}
where $R_0$ is the resistance in the limit of zero current and $\alpha$ is the  parameter describing the nonreciprocity in transport. We neglect higher order terms in $I$. For such a system, an a.c. current at frequency $f$ with r.m.s. amplitude $I_{ac}$ results in a r.m.s. voltage $V_{2f}=\frac{1}{\sqrt{2}}\alpha I_{ac}^2$ at the second harmonic frequency $2f$. (See Supplemental Material for the derivation). An a.c. current ($I_{ac}$ = 5 $\mu$A, $f$ = 13.27 Hz) was applied between contacts C1 and C2 of sample D2. $V_{2f}$ was measured between probes U1 and U4. The corresponding resistance $R_{2f} = V_{2f}/I_{ac}$ is shown in Fig. 2e. We observed sharp peaks in $R_{2f}$ at the critical field transition (determined by resistance measurements in Fig. 2c). This shows that the superconducting transition is indeed associated with even-in-field nonreciprocal transport properties and a finite value of the nonreciprocity parameter $\alpha$.

The resistance of the conduction channel between the contacts U7 and U8 was measured in two-probe configuration as a function of temperature (Fig. 2f). The narrow segments (marked by a dashed ellipse in Fig. 2b) have a broad transition with a $T_c$ of 2.5 K. The zero resistance state is still not reached at the lowest accessible temperature of 2.0 K. The critical magnetic field of these segments at 2.0 K was determined to be 0.6 T from resistance measurements (Fig. 2g). Fig. 2h shows the measurement of spontaneous voltage without an applied current, using a Keithley 2182A nanovoltmeter. The transitions of wide and narrow regions correspond to two voltage peaks around 4.0 T and 0.6 T respectively. The peaks arising at the superconducting transition of the narrow regions show step-like features. A spontaneous voltage was also seen to develop at the superconducting transition induced by varying the temperature at zero magnetic field (Fig. 2i). It is therefore evident that the phenomenon of rectification is associated with both the field-induced and temperature-induced transitions. If the temperature is held constant, the voltage is stable over time (Fig. 2j).

We carried out voltage measurements across the probes U7 and U8 by interchanging the polarities of the nanovoltmeter leads (Fig. 2k). In both cases, there were positive peaks. The features of the voltage curves vary in their details from one set of measurement to another, as can be seen in Fig. 2k, and also in comparison with Fig. 2h. The general feature is that the peaks always occur across the broad transition seen in resistivity measurements (Fig. 2g). We now turn to the discussion of the observation that the voltage peaks do not change sign upon interchanging the voltmeter input connectors. Voltmeters used in practice are not ideal instruments and may apply small signals on the sample being measured. One such signal is the input bias current \cite{keithley} which flows between its terminals. This effect is negligible under most practical circumstances. The nanovoltmeter bias current is typically 60 pA, producing a drop of 60 nV per k$\Omega$ of external resistance. Other sources of signals from the voltmeter might be stray voltages due to charges accumulated across the input impedance and thermoelectric potentials in its internal circuitry. We speculate that it is precisely such small and undetermined signals which cause the polarity of the spontaneous d.c. voltage peak across the sample to develop in a specific orientation, in correlation with the polarity of connectors at the voltmeter input. To verify this understanding, the circuit outlined in Fig. 2l was implemented. Here, the device (with resistance $R_s$) is connected in parallel with a resistance $R_p$. When $R_p \ll R_s$, it lessens the impact of any signal originating within the voltmeter on the sample. The spontaneous voltage which develops across the sample changes drastically, because in order to maintain the same voltage drop across a small $R_p$, a far greater amount of power needs to be delivered which places severe constraints on the physical system. Despite this drawback, this set-up enables us to test whether lowering the impact of signals from the voltmeter allows the polarity of spontaneous voltage to develop independently of the orientation of the voltmeter leads, or not. Upon conducting these measurements (Fig. 2l) with $R_p=$ 67 $\Omega$, we found that the measured voltage features indeed reversed the sign upon interchanging the voltmeter leads. This shows that the usual observation of positive voltage peaks (when no $R_p$ is present) is likely to be a consequence of small signals  within the voltmeter. These observations highlight that the polarity of spontaneous voltage at the superconducting transition is extremely susceptible to small external fields. It indicates that a mechanism of spontaneous symmetry breaking might be at work. An analogy may be drawn to the three dimensional Ising model of spins \cite{reif, goldenfeld} with attractive interaction, where the spins are allowed to point either `up' or `down'. As is well known, a net magnetization $M$ develops below a critical temperature due to spontaneous breaking of up-down symmetry. The preferred direction (`up' or `down') may be determined by an infinitesimally small external magnetic field. In our experiments, the d.c. voltage seemingly has its origin in rectification effects due to a spontaneous breaking of inversion symmetry along the Nb channel at the superconducting transition. A small external electric field is capable of inducing a preferred direction, which in our measurements is provided by the measuring instrument (voltmeter).

From our empirical results, we postulate that the rectification effect seen here is associated with the spontaneous development of an electric polarization across the length of the sample, resulting in a potential landscape without inversion symmetry. The origin of such a polarization should be related to the appearance of dipole moments at the microscopic scale. We will now discuss a possible mechanism of how such microscopic dipoles can arise. For this, we will focus on two components of the niobium film - the system of conducting electrons within which superconductivity develops, and non-metallic niobium grains which do not contribute to superconductivity.

Moro et al. \cite{moro} reported the signature of ferroelectricity in clusters of Nb atoms. These dipoles are nonclassical in nature \cite{xu} and can not be understood in terms of a polarization built into the structure of the cluster. The dipole moments in such clusters bear the signature of a spontaneous-symmetry-broken state. Durkin et al. \cite{durkin} studied the transport properties of niobium nanoislands and concluded that superconductivity develops from a mechanism of rare-region onset with only very large grains contributing to superconducting order. Bose et al. \cite{bose} observed that superconductivity disappeared for niobium films with an average particle size of 8 nm, consistent with the Anderson criterion. At low temperatures, insulating behaviour might be expected in very small grains which are a few nanometres in size \cite{bose2}, as well as in intergrain boundaries. We hypothesize that the insulating parts of the niobium films host nonclassical dipole moments similar to those observed in the clusters in Refs. \cite{moro} and \cite{xu}. The mutual interaction between these dipole moments in insulating grains leads to the development of a macroscopic electric polarization $\mathbf{P_{i}}$. We now turn our attention to the system of conducting electrons which arises from the larger and more metallic grains and in which superconductivity develops. Theoretical studies \cite{khomskii, khomskii2} have shown that the chemical potentials of superconducting and normal states are different, as a consequence of which there is a transfer of charge from one electronic subsystem to another if they coexist. Therefore, the presence of two different phases in the vicinity of the superconducting transition leads to the development of dipole moments at the interface of superconducting and normal regions within the niobium film. The total polarization $\mathbf{P_{el}}$ within the electronic system, on taking the macroscopic average, should generally be zero. However, when there exists a macroscopic electric polarization $\mathbf{P_{i}}$ in the background originating from the insulating grains, $\mathbf{P_{el}}$ can have a finite non-zero value. Conduction electrons experience the electric field due to the polarization $\mathbf{P_{el}}$ at the interface of superconducting and normal regions. This offers a possible mechanism of the asymmetry leading to a rectification effect.

In summary, we have observed spontaneous voltage peaks at the superconducting transition of plain Nb devices resulting from a rectification of environmental fluctuations. The key observations are that the peaks are symmetric in the magnetic field and are present at the critical temperature without an applied magnetic field. This signifies an unconventional mechanism of rectification which is in stark contrast to the commonly observed antisymmetric field-dependence of the rectified signal. The spontaneously arising asymmetry in our devices is possibly related to the development of an electric polarization at the superconducting transition. We can account for this phenomenon based on the hypothesis that insulating regions of niobium films (nanosized grains and intergrain boundaries) host non-classical dipole moments. Experimental \cite{tao, glover, desimoni} and theoretical \cite{hirsch, hirsch2} works have previously suggested that there are many interesting questions regarding the impact of electric fields on superconductors which are yet to be fully understood. Our experiments show that rectification effects can reveal the existence of electric polarization concomitant with superconductivity and provide new directions for research on these questions.

\section{Acknowledgments}
We thank Jana Lustikova, Alejandro Silhanek and Xiaoshan Xu for insightful discussions. We are grateful to Laurent Berg\'{e} and Sophie Gu\'{e}ron for help with fabrication of devices. This work was supported by public grants from the French National Research Agency (ANR), project CP-Insulators No. ANR-2019-CE30-0014-03.

\widetext
\newpage
\begin{center}
\textbf{\large Supplemental Material}
\end{center}
\setcounter{equation}{0}
\setcounter{figure}{0}
\setcounter{table}{0}
\setcounter{page}{1}
\makeatletter
\renewcommand{\theequation}{S\arabic{equation}}
\renewcommand{\thefigure}{S\arabic{figure}}
\renewcommand{\bibnumfmt}[1]{[S#1]}
\renewcommand{\citenumfont}[1]{S#1}

\bigskip

\bigskip

\bigskip

\bigskip

\begin{flushleft}

\textbf{1. Second harmonic component in voltage with an a.c. bias current}

\bigskip

The simplest expression to describe a directional anisotropy in resistance (\textit{R}) as a function of the current (\textit{I}) is
\end{flushleft}
\begin{equation}
R=R_0+\alpha I,
\end{equation}

\begin{flushleft}

where $R_0$ is the resistance in the limit of zero current and $\alpha$ is the coefficient describing the anisotropy. Let us consider the case of an a.c. current of amplitude $\widetilde{I_{ac}}$ at frequency $f$ applied through the system. 
\end{flushleft}
\begin{equation}
I=\widetilde{I_{ac}} sin(2\pi ft),
\end{equation}
with \textit{t} denoting time.
\begin{flushleft}

The corresponding voltage drop ($V=RI$) is:
\end{flushleft}
\begin{equation}
V=\frac{1}{2}\alpha{\widetilde{I_{ac}}}^2 + R_0\widetilde{I_{ac}} sin(2\pi ft) + \frac{1}{2}\alpha {\widetilde{I_{ac}}}^2 sin (2\pi.2ft-\pi/2)
\end{equation}

\begin{flushleft}

Expressed in terms of the r.m.s. current $I_{ac}=\widetilde{I_{ac}}/\sqrt{2}$, the r.m.s. voltage $V_{2f}$ at the second harmonic frequency ($2f$) is
\end{flushleft}
\begin{equation}
V_{2f}=\frac{1}{\sqrt{2}}\alpha {I_{ac}}^2
\end{equation}

\bigskip

\bigskip

\begin{flushleft}

\textbf{2. Characterization of sample D1}

\bigskip

Fig. S1a shows the image of sample D1 observed with a scanning electron microscope. The conduction channel between the two contact pads has a length of 245 $\mu$m and a width of 38 $\mu$m. The topography of the edge of this channel, measured with an atomic force microscope, is displayed in Fig. S1b. The thickness of the Nb film is 72 nm.

\bigskip

\bigskip

\newpage

\textbf{3. Transport measurements on sample D2}

\bigskip

Sample D2 consists of a long conduction channel of length 1380 $\mu$m connected to the contacts pads marked as C1 and C2 (Fig. 2a of the main text). The voltage probes have a width of only a few micrometres near the junction with the central conduction channel (Fig. 2b of main text). One such junction was scanned with an atomic force microscope (Fig. S2). The thickness of the film is 55 nm.

\begin{figure}[h]
\begin{center}
\includegraphics[width=140mm]{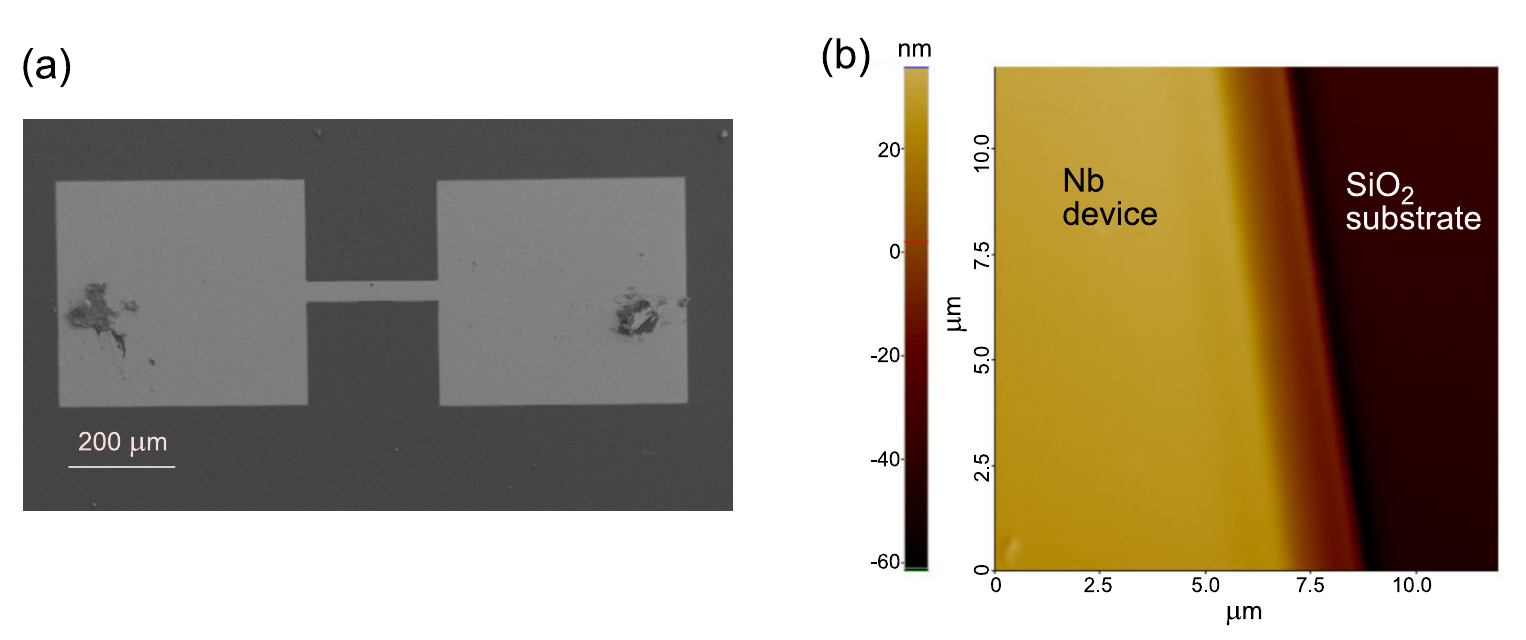}
\caption{\textbf{(a)} Scanning electron microscope image of sample D1. \textbf{(b)} Image of one edge of the conduction channel observed with an atomic force microscope.}
\end{center}
\end{figure}

\begin{figure}[h]
\begin{center}
\includegraphics[width=60mm]{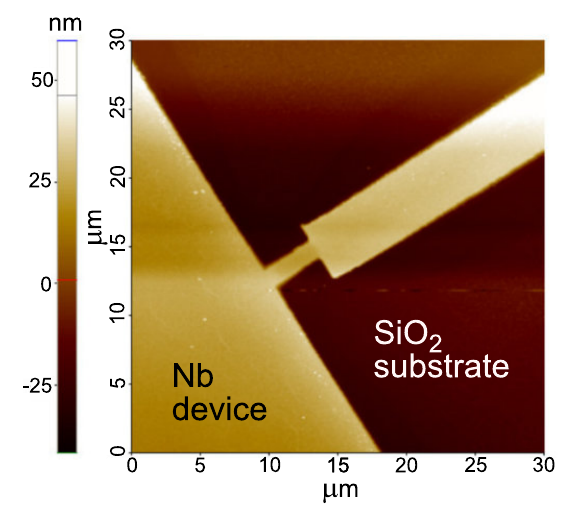}
\caption{Image of sample D2 with an atomic force microscope.}
\end{center}
\end{figure}

\bigskip

\begin{figure}[h]
\begin{center}
\includegraphics[width=160mm]{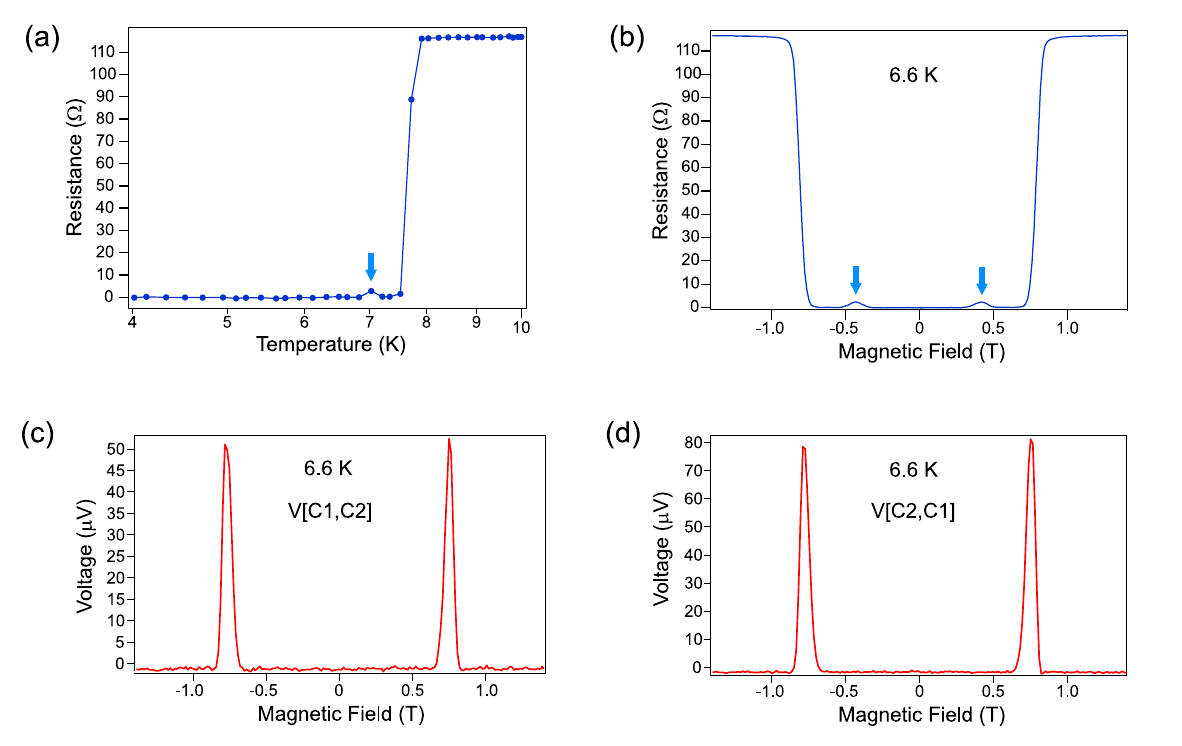}
\caption{\textbf{(a)} Four-probe measurement of the critical superconducting temperature in sample D2 using a d.c. current bias of 1 $\mu$A. The current was applied between contact pads C1 and C2. (See Fig. 2a of main text for the nomenclature of contacts.) The voltage drop was measured between U1 and U4 with a DMM. \textbf{(b)} Measurement of the critical magnetic field at a temperature of 6.6 K in the same setup. The d.c. current applied was 5 $\mu$A. \textbf{(c,d)} Spontaneous voltage measured on contacts C1 and C2 at 6.6 K using a DMM. The two plots show the data for two different configurations of the voltmeter leads, as outlined in Fig. 1d of the main text.}
\end{center}
\end{figure}

\begin{figure}[h]
\begin{center}
\includegraphics[width=160mm]{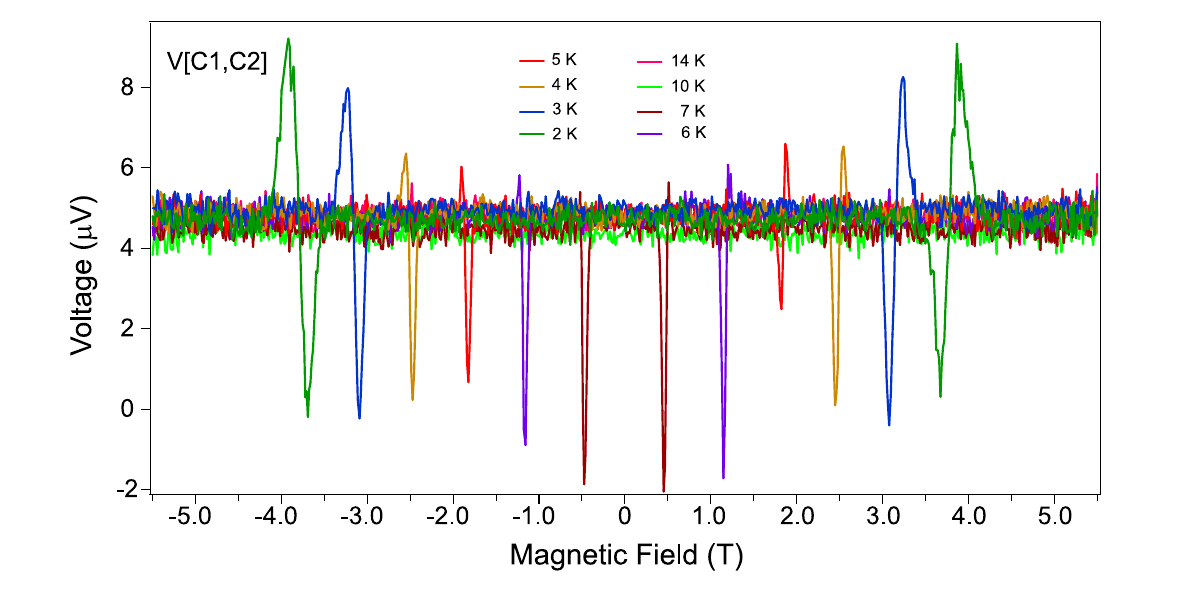}
\caption{Spontaneous voltage measured between contacts C1 and C2 using a DMM while sweeping the magnetic field.}
\end{center}
\end{figure}

\begin{figure}[h]
\begin{center}
\includegraphics[width=164mm]{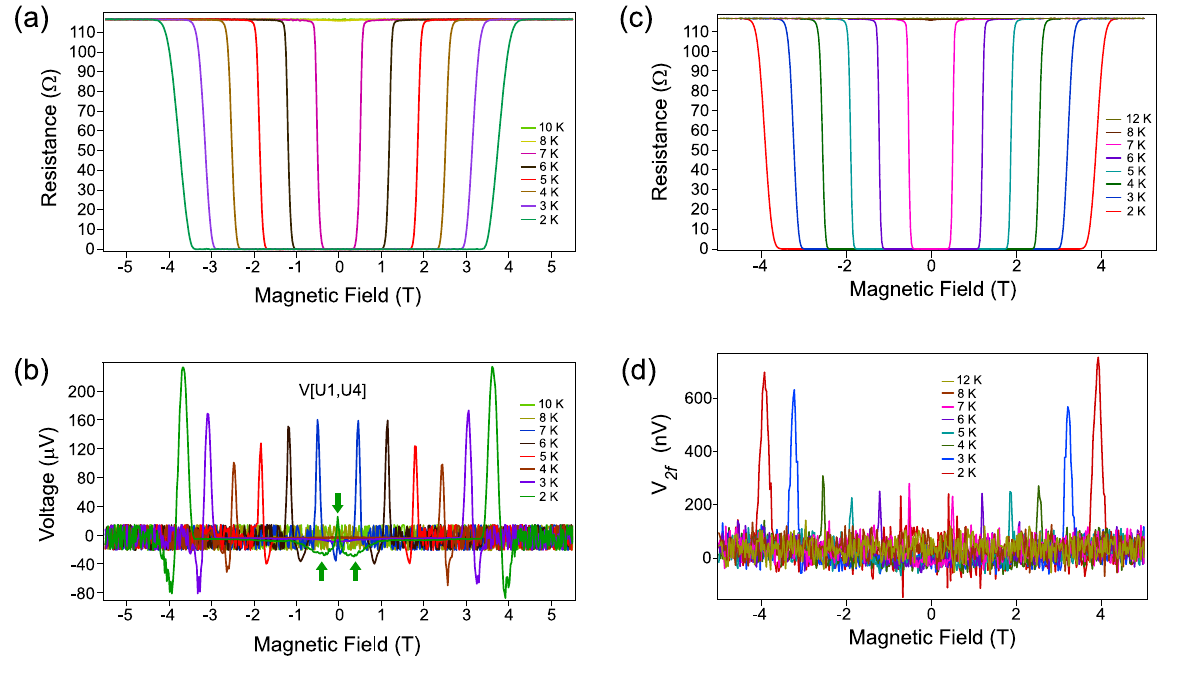}
\caption{\textbf{(a,b)} Simultaneous measurement of resistance in four-probe configuration and spontaneous d.c. voltage. An a.c. current of r.m.s. amplitude 1 $\mu$A and frequency ($f$) of 13.27 Hz was applied between the contacts C1 and C2. The voltage drop was measured between U1 and U4 with a lock-in amplifier. The resistance thus measured is shown in (a). The d.c. voltage, displayed in (b), was measured between the same contacts (U1 and U4) with a DMM. \textbf{(c,d)} Simultaneous measurement of resistance and second harmonic voltage in four-probe configuration. An a.c. current of r.m.s. amplitude 5 $\mu$A ($f$ = 13.27 Hz) was applied between the contacts C1 and C2. One lock-in amplifier measured the voltage between U1 and U4. The resistance thus calculated is shown in (c). A second lock-in ampifier measured the second harmonic voltage $V_{2f}$, shown in (d).}
\end{center}
\end{figure}

\begin{figure}[h]
\begin{center}
\includegraphics[width=100mm]{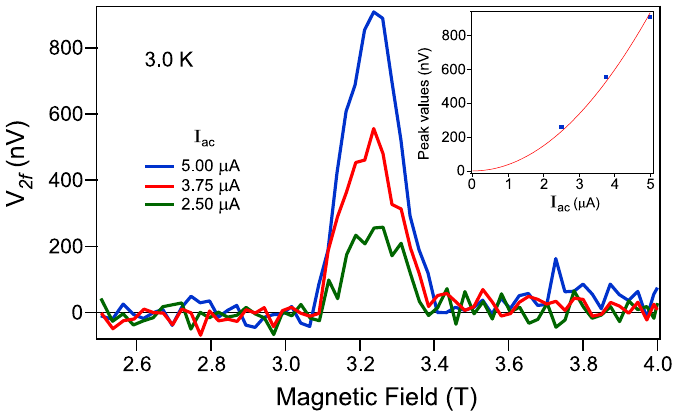}
\caption{Measurement of $V_{2f}$ for different a.c. bias currents. The a.c. current of r.m.s. amplitude $I_{ac}$ (\textit{f}=13.27 Hz) was applied between the contacts C1 and C2. The voltage $V_{2f}$ was measured across U1 and U4.}
\end{center}
\end{figure}

\begin{figure}[h]
\begin{center}
\includegraphics[width=140mm]{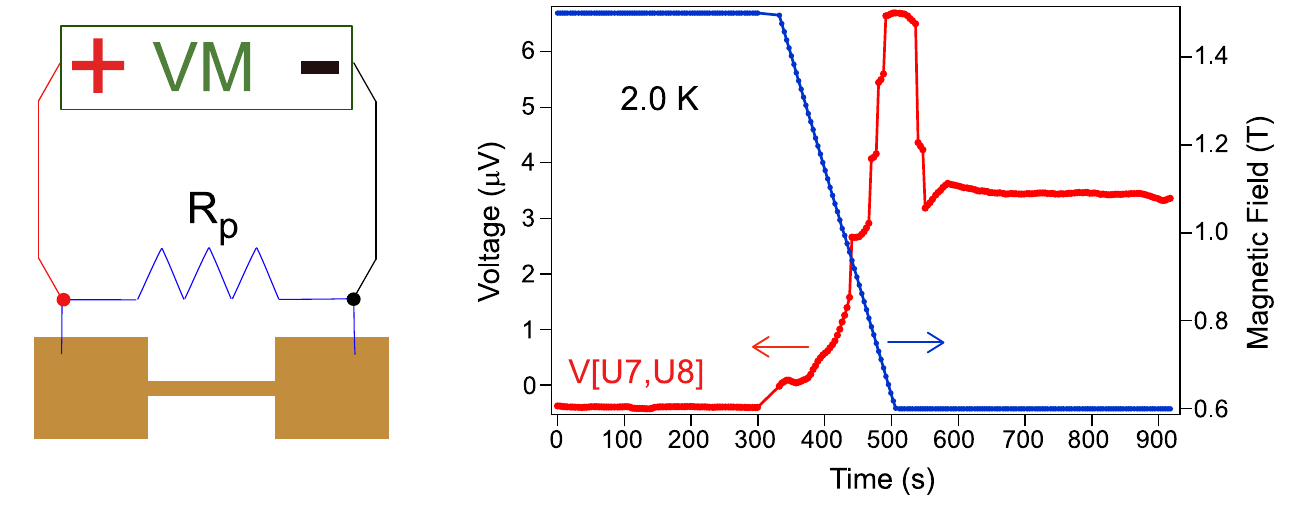}
\caption{Measurement of spontaneous d.c. voltage with a 10 k$\Omega$ external resistor in parallel. A resistance $R_p$ of value 10 k$\Omega$ was put across the connectors (at room temperature outside the cryostat) leading to the contacts U7 and U8 on the sample. The voltage across these connectors was measured with a nanovoltmeter while sweeping the magnetic field. The temperature of the Nb device in the cryostat was 2.0 K. The magnetic field was held constant when it reached 0.600 T and the voltage was recorded with time.}
\end{center}
\end{figure}

In Fig S3a, we show four probe measurements on the device. The d.c. current (1 $\mu$A) was applied between C1 and C2. The voltage drop was measured between the probes U1 and U4 using a digital multimeter (DMM). The superconducting transition occurs at 7.7 K. There is a feature of a small peak (indicated by an arrow in Fig. S3a) at 7.0 K. A similar feature was also visible (at a magnetic field of 420 mT) when the superconducting transition was induced by varying the magnetic field (Fig. S3b) at 6.6 K. The critical field of the central conduction channel (Fig. S3b) at 6.6 K is 800 mT. The feature at 420 mT (indicated by arrows) possibly arises from rectification of environmental electromagnetic fluctuations within the voltage probes. The characterization of voltage probes (using contacts U7 and U8), shown in Fig. 2f of main text, reveals a transition at 7.0 K. The small peak feature in Fig. S3a occurs at a similar temperature, suggesting that the cause may be attributed to rectification along the voltage probes. Ideally, electrical measurements in the four-probe configuration are not supposed to be affected by the properties of the voltage probes. It is interesting to note that in the presence of rectification effects, this ideal scenario is not respected. We therefore need to be cautious while interpreting the results of transport experiments on superconducting devices, since such features arising within voltage probes may add unexpected voltage contributions to the measured signal.

\bigskip

Magnetotransport measurements with a current applied along the central channel of sample D2 reveal that its mean free path ($l$), Fermi wave vector ($k_F$) and superconducting coherence length ($\xi$) are 3.2 nm, 11.2 nm$^{-1}$ and 9.1 nm respectively. 

\bigskip

Figs. S3c and S3d show the measurement of spontaneous voltage between contacts C1 and C2 with a DMM at 6.6 K. The peaks are positive, as we have observed on most instances. We generally observe that the peak heights show some differences from one experimental run to another. When measurements are performed on different days, the magnitude of the peaks can also be quite different. We attribute these differences to an overall change in the amount of environmental fluctuations, which is not a controlled parameter. Some difference can also arise from the behaviour of the sample itself.  At a microscopic level, the spontaneously arising asymmetry in the sample is possibly related to the redistribution of charges owing to the presence of superconducting and non-superconducting domains. The configuration of these domains are likely to be different from one set of measurement to another.

\bigskip

Although in most instances we observe positive peaks, we have observed negative peaks a few times. One dataset is shown in Fig. S4. The peaks are always symmetric in the magnetic field.

\bigskip

Figs. 2c and 2d of the main text show the four-probe resistance (measured with a lock-in technique) and the spontaneous d.c. voltage in sample D2. The complete dataset is shown in Figs. S5a and S5b. At 2 K (Fig. S5b), there are features below 1 T (indicated by arrows). These arise from the narrow channels of the voltage probes (U1 and U4). We do not see these features above 3 K, since these temperatures are above the critical temperature for these narrow channels.

\bigskip

In Fig. 2e of the main text, the data for measurement of second harmonic voltage was displayed. For these measurements, we used two lock-in amplifiers measuring the voltage on contacts U1 and U4. One lock-in amplifier measured the voltage at the same frequency while the other measured the voltage $V_{2f}$ at the second harmonic frequency. The complete dataset is presented in Figs. S5c and S5d.

\bigskip

It is predicted that the second harmonic voltage $V_{2f}$ will vary as $I_{ac}^2$ (Eq. S4). We performed measurements of $V_{2f}$ for three different values of $I_{ac}$ at 3 K (Fig. S6). The peak heights are plotted in the inset of Fig. S6. These indeed follow the $I_{ac}^2$ dependence, with the fit parameter $\alpha$ = 5.2 $\times$ $10^4$ $\Omega$/A.  An underlying assumption during this analysis is that the anisotropy parameter $\alpha$ is a constant. This is found to be generally true when the measurements are done in quick succession, as in this case. However, when performed in a different experimental run, the results can be different. This is seen when comparing the results of Fig. S6 (for $I_{ac}$ = 5 $\mu$A) with the plot in Fig. S5d (for a temperature of 3 K). The experimental setup was identical in both cases, but the peak heights are not the same.

\bigskip

It was discussed in the main text that the measurement of a d.c. voltage in a voltmeter implies that work is being done across its internal impedance. This consideration led us to conclude that the spontaneous voltage arises from the rectification of environmental fluctuations. The fact that work is being done can be demonstrated more clearly by adding a known resistor in parallel with the sample. We measured the voltage across the probes U7 and U8 with a 10 k$\Omega$ resistor in parallel. The field was ramped till 600 mT (upper critical field of superconductivity) and then held constant, with the temperature of the sample at 2.0 K. With the magnetic field set at 600 mT, the observed voltage was steady over several minutes (Fig. S7) at approximately 3.4 $\mu$V. This voltage corresponds to a steady power dissipation of 1.2 fW across the 10 k$\Omega$ resistor.

\bigskip

\bigskip

\bigskip

\newpage

\textbf{4. Generation of d.c. voltage due to mixing of d.c. and a.c. signals}

\bigskip

The phenomenon of non-reciprocal transport is at the heart of the rectification effect seen in our experiments. However, the generation of a d.c. voltage from a.c. fluctuations can occur under a very different scenario as well, which does not require the presence of any directional asymmetry in the system. This may happen in isotropic systems with non-linear voltage-current relation, for which the resistance ($R$) can be expressed as an expansion in even powers of the current $I$. 
\end{flushleft}
\begin{equation}
R = R_0 +\sum_{i=2,4,6...}^{\infty}\eta_i I^i
\end{equation}
\begin{flushleft}
In the above expression, $R_0$ is the resistance in the limit of zero current and $\eta_i$ denotes the coefficient of the term $I^i$ ($i$ being an even integer). It follows that the voltage $V$=$RI$ is an odd function of $I$. A simple form of Eq. S5 with the lowest order term in $I$ is:
\end{flushleft}
\begin{equation}
R=R_0+\eta_2 I^2.
\end{equation}
\begin{flushleft}

\bigskip

Let us consider a situation where a d.c. current $I_{dc}$ is applied along with an a.c. current of amplitude $\widetilde{I_{ac}}$ and frequency $f$.
\end{flushleft}
\begin{equation}
I=I_{dc}+\widetilde{I_{ac}} sin(2\pi ft)
\end{equation}

\begin{flushleft}

Since the voltage $V=RI$ involves the term $\eta_2 I^3$, the a.c. current contributes to the d.c. voltage $V_{dc}$ and generates a voltage $\widetilde{V_{2f}}$ at frequency $2f$. 
\end{flushleft}
\begin{equation}
V_{dc} = R_0I_{dc} + \eta_2 I_{dc}^3 + \frac{3}{2} \eta_2 I_{dc} \widetilde{I_{ac}}^2
\end{equation}
\begin{equation}
\widetilde{V_{2f}} = \frac{3}{2} \eta_2 I_{dc} \widetilde{I_{ac}}^2 sin(2\pi.2ft - \frac{\pi}{2})
\end{equation}

\begin{flushleft}

The above equations show that, for an a.c. current passing through a resistance described by Eq. S6, the generation of a d.c. voltage as well as an a.c. voltage at second harmonic frequency is indeed possible. Similar results can also be derived starting from the more general expression Eq. S5. It can be shown that each term involving $I^i$ leads to the generation of d.c. voltage contributions with two terms varying as ${I_{dc}}^{i+1}$ and ${I_{dc}}^{i-1}$ ${\widetilde{I_{ac}}}^2$ (assuming ${\widetilde{I_{ac}}}\ll{I_{dc}}$).

\bigskip

Unlike the case of non-reciprocal transport (described by Eq. S3), $V_{dc}$ and $\widetilde{V_{2f}}$ calculated above (Eqs. S8 and S9) have finite values only when there is a d.c. current $I_{dc}$ in the system. The presence of a voltmeter connected to the device implies that the input bias current ($I_B$) of the instrument is passing through the sample. Further, we can expect small non-linearities in voltage-current relation (finite $\eta_i$) at the superconducting transition. Therefore, it may be argued that all the necessary conditions for the generation of a d.c. voltage from electromagnetic fluctuations (as described in Eq. S8) are satisfied in our experiments. This implies a trivial explanation of the spontaneous voltage peaks observed in our devices without any requirement for invoking non-reciprocal transport. We present below results to show that this is actually not the case.

\bigskip

Fig. S8a shows a diagram of the circuit implemented to measure voltage-current (\textit{V-I}) characteristics across the probes U7 and U8 of sample D2. The voltage across the two contacts was measured with a nanovoltmeter. This instument has $I_B$ $<$ 60 pA. An external current was applied beween the same contacts, as shown by the current source $I_S$ is Fig. S8a. If we assume that the d.c. voltage peaks at the superconducting transition result from the mixing of d.c. and a.c. signals (Eq. S8), with $I_{dc}$ being the same as $I_B$, we can expect a reduction of the generated $V_{dc}$ to zero when $I_B$ is offset by an externally applied current in the opposite direction. 

\bigskip
Fig. S8b shows the \textit{V-I} characteristic at a temperature of 2.0 K and applied magnetic field of 1.8 T. At this value of the magnetic field, the narrow channels along the path connecting U7 and U8 are close to the onset of the superconducting transition (see Fig. 2g of the main text). The peak  of the spontaneous voltage is expected to develop at much lower fields (Fig. 2h of main text). The linear slope of the \textit{V-I} curve (Fig. S8b) corresponds to a resistance of 940 $\Omega$.

\bigskip
Following the above measurement, the magnetic field was further reduced without any externally applied current to induce the superconducting transition in the device. As expected, a spontaneous voltage was seen to develop. The magnetic field sweep was stopped at 0.6 T. The spontaneous voltage ($V_0$) observed was 7 $\mu$V. The \textit{V-I} curve measured at this value of the magnetic field is displayed in Fig. S8c. The applied current spanned a wide range from -1 nA to 1 nA. This is much larger than the input bias current of the nanovoltmeter. The voltage changed very little and was far from approaching zero. This result demonstrates that the spontaneous voltage observed in our devices is not arising from the mixing of d.c. and a.c. signals as described in Eq. S8. 

\bigskip

While sweeping the d.c. current, it was found that the voltage measured was more noisy for faster rates of current sweep. The data of Fig. S8c was acquired at a sweep rate of 0.8 pA/s. Fig. S8d shows the measurement of \textit{V-I} characteristics over a wider current range at a faster sweep rate of 28 pA/s. A linear fit to the data (Fig. S8d) reveals a resistance of 555 $\Omega$. This value is consistent with the resistance measurements shown in Fig. 2g of the main text in the same device, where a d.c. current of 78 nA was used. It confirms that the evolution of $V_{dc}$ observed in the \textit{V-I} measurements occurs entirely due to the resistive voltage drop ($R_0I_{dc}$) and does not involve the mixing of the voltmeter input bias current with environmental electromagnetic fluctuations.

\bigskip

\begin{figure}[h]
\begin{center}
\includegraphics[width=160mm]{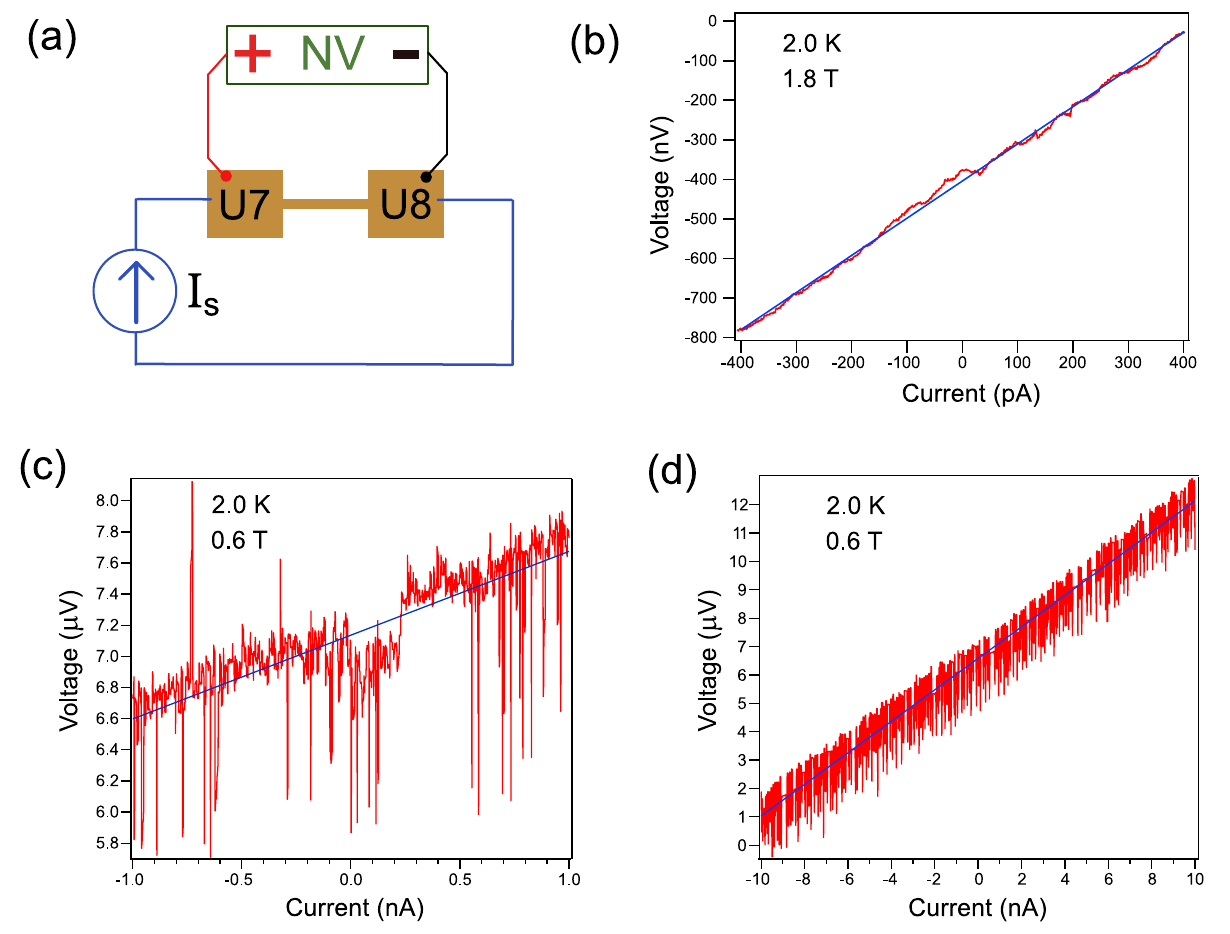}
\caption{\textbf{(a)} Circuit implemented to measure voltage-current (\textit{V-I}) characteristics across the probes U7 and U8 of sample D2. \textbf{(b)} \textit{V-I} curve measured at 2.0 K temperature with an applied magnetic field of 1.8 T. The linear fit gives a resistance of 940 $\Omega$. \textbf{(c,d)} \textit{V-I} curves at an applied magnetic field of 0.6 T. Linear fits to the data correspond to resistance values of 538 $\Omega$ (c) and 555 $\Omega$ (d). The voltage at zero current obtained from the fit is 7.1 $\mu$V in (c) and 6.6 $\mu$V in (d).}
\end{center}
\end{figure}

\end{flushleft}

\end{document}